# Analytic model for the surface potential and drain current in negative capacitance field-effect transistors

David Jiménez and Enrique Miranda
Departament d'Enginyeria Electrònica, Escola d'Enginyeria, Universitat Autònoma de Barcelona, 08193-Bellaterra, Barcelona, Spain

Andrés Godoy
Departamento de Electrónica y Tecnología de Computadores, Facultad de Ciencias, Campus de Fuente Nueva s/n, Universidad de Granada, Granada, Spain

Corresponding author: david.jimenez@uab.es

*Abstract*.- **In 2008, Salahuddin and Datta [1] proposed that a ferroelectric material operating in the negative capacitance region could act as a step-up converter of the surface potential in a MOS structure, opening a new route for the realization of transistors with steeper subthreshold characteristics (*S*<60 mV/decade). In this letter, a comprehensive physics-based surface potential and drain current model for the negative capacitance field-effect transistor is reported. The model is aimed to evaluate the potentiality of such transistors for low-power switching applications. Moreover it provides a model core for memories devices relying on the hysteretic behavior of the ferroelectric gate insulator.**

**Introduction**

One of the most severe problems pointed out in the *International Technology Roadmap for Semiconductors* [2] is the increasing power dissipation density at the chip level caused by the relentless scaling down of transistors. This problem is often referred to as the power crisis in the microelectronics industry and its root can be traced back to the difficulty in increasing the transistor subthreshold slope, which, because of thermodynamic considerations, exhibits an apparent limit of $S$=60 mV/decade at room temperature, where $S$ is an inverse slope. This apparent limitation has been termed as the Boltzmann limit. It has been recently suggested [1] that a metal-ferroelectric-semiconductor (MFS) gate stack could provide a new mechanism to step-up the semiconductor surface potential ($\varphi_s$) above the gate voltage. The underlying idea consists in exploiting the negative capacitance region of the ferroelectric material (Fig. 1a; red dashed line), defined as $C_f$=d$Q$/d$V_f$, where $Q$ and $V_f$ refers to the charge density per unit area and the voltage drop in the ferroelectric, respectively. Provided the MFS gate stack could be operated in the region where $C_f$<0, the surface potential $\varphi_s$, arising from the capacitance divider formed by $C_f$ and the semiconductor capacitance, could be up-converted. This phenomenon could be used to enhance the electrostatic control of the gate over the channel of a field-effect transistor (FET), thus opening a route towards steeper subthreshold transistors exhibiting $S$<60 mV/decade. The negative capacitance FET proof of concept has not been demonstrated yet, and many open issues remain to be solved. At the time being, the main challenge is to identify a suitable MFS gate stack able to operate in the negative capacitance region with the consequent step-up conversion of the surface potential. A promising way involves the use of organic ferroelectrics considered by Ionescu's group [3]. In parallel with efforts to proof the concept, other developments have to be made in order to adequately model the surface potential up-conversion and how this effect impacts on the drain current. To fill this gap, an analytic physics-based surface potential and drain current model for the negative capacitance FET valid for all operation regions is proposed. For the sake of simplicity, a double-gate (DG) geometry was assumed, although the model can be easily extended to other geometries such as for instance, single-gate or surrounding-gate geometries [4-6].

**Model**

Consider an undoped or lightly doped, negative capacitance FET with a symmetric DG geometry (Fig. 1b). According to the gradual channel approximation, Poisson's equation takes the following form along a vertical cut perpendicular to the semiconductor film:

$$\frac{d^2\varphi}{dx^2} = \frac{q}{\varepsilon_s} n_i e^{\frac{q(\varphi-V)}{kT}} \quad (1)$$

where $q$ is the electron charge, $n_i$ the intrinsic carrier concentration, $\varepsilon_S$ the permittivity of the semiconductor, $\varphi(x)$ the electrostatic potential, and $V$ the electron quasi-Fermi potential. Notice that the right side of (1) only contains the free charge concentration. This simplification is valid for q$\varphi$/kT>>1, being the hole density negligible. Equation (1) must satisfy the following boundary conditions:

$$\frac{d}{dx}(x=0) = 0, \qquad (x = \pm\frac{t_s}{2}) = \varphi_s \quad (2)$$

where $t_s$ is the semiconductor film thickness. Since the current mainly flows along the y-direction, we assume that $V$ is constant along the x-direction invoking the gradual channel approximation; i.e., $V=V(y)$. The analytical solution for the electrostatic potential in the semiconductor can be written as [7]:

$$\varphi(x) = V - \frac{2kT}{q} \ln\left[\frac{t_s}{2\beta}\sqrt{\frac{q^2 n_i}{2\varepsilon_s kT}} \cos\left(\frac{2\beta x}{t_s}\right)\right] \quad (3)$$

where $\beta$ is a constant, to be determined from the boundary condition

$$V_g - \Delta\phi - \varphi_s = a_0 Q + b_0 Q^3 + c_0 Q^5 \quad (4)$$

where $V_g$ is the voltage applied to both gates and $\Delta\phi$ the work-function difference between the gate electrode and the intrinsic semiconductor. Note that the left side of (4) corresponds to $V_f$. The right side is representative of the $Q$-$V_f$ characteristic of a ferroelectric oxide given by the phenomenological Landau-Ginzburg-Devonshire theory, although a higher degree polynomial could be needed to properly describe the $Q$-$V_f$ characteristic. The coefficients $a_0$, $b_0$, $c_0$ are related to the Landau parameters $a$, $b$, $c$, of the ferroelectric material by the following relationships [8]: $a_0 = 2t_f a$, $b_0 = 4t_f b$, $c_0 = 6t_f c$, where $t_f$ is the ferroelectric film thickness. The coefficient $a_0 \sim 1/C_f$ represents the inverse of the ferroelectric capacitance at low $V_f$, and it is assumed to be $a_0 < 0$ as the key feature for up-conversion. From Gauss's law, the total mobile charge per unit gate area can be determined by $Q=2\varepsilon_s(d\varphi(t_s/2)/dx)$, which equals $(2\varepsilon_s)(2kT/q)(2\beta/t_s)\tan\beta$. Substituting $Q(\beta)$ into (4) leads to:

$$\frac{q(V_g - \Delta\varphi - V)}{2kT} - \ln\left(\frac{2}{t_s}\sqrt{\frac{2\varepsilon_s kT}{q^2 n_i}}\right) = \ln(\beta) - \ln(\cos\beta) + a_0(2C_s)\beta\tan\beta +$$
$$+ b_0(2C_s)^3\left(\frac{4kT}{q}\right)^2 \beta^3 \tan^3\beta + c_0(2C_s)^5\left(\frac{4kT}{q}\right)^4 \beta^5 \tan^5\beta \quad (5)$$

where $C_s = \varepsilon_s/t_s$ is a structural capacitance. Importantly, (5) reduces to Taur's expression for a DG-MOSFET after the identification $a_0 = 1/C_{ox}$, $b_0 = c_0 = 0$, where $C_{ox}$ represents the oxide capacitance. For a given $V_g$, $\beta$ can be solved from (5) as a function of $V$. Note that $V$ varies from source to drain. The functional dependence of $V(y)$ and $\beta(y)$ are determined by the current continuity equation, which requires the (drift-diffussion) current $I_{ds}=\mu WQdV/dy$=constant, independent of $V$ or $y$. The parameter $\mu$ is the effective mobility in the semiconductor and $W$ the device width. Integrating $I_{ds}dy$ from the source to the drain and expressing $dV/dy$ as $(dV/d\beta)(d\beta/dy)$, the current can be written as

$$I_{ds} = \mu \frac{W}{L} \int_0^{V_{ds}} Q(V)dV = \mu \frac{W}{L} \int_{\beta_s}^{\beta_d} Q(\beta) \frac{dV}{d\beta} d\beta \quad (6)$$

where $\beta_s$, $\beta_d$ are solutions to (5) corresponding to the cases $V=0$ and $V=V_{ds}$ respectively. Note that $dV/d\beta$ can also be expressed as a function of $\beta$ by differentiating (5). Substituting these factors in (6), integration can be performed analytically to yield:

$$I_{ds} = \mu \frac{4C_s W}{L}\left(\frac{2kT}{q}\right)^2 \int_{\beta_S}^{\beta_D} \left[\tan\beta + \beta\tan^2\beta + a_0(2C_s)\beta\tan\beta\frac{d}{d\beta}(\beta\tan\beta) + \right.$$

$$+ 3b_0(2C_s)^3\left(\frac{4kT}{q}\right)^2(\beta\tan\beta)^3\frac{d}{d\beta}(\beta\tan\beta) +$$

$$\left. + 5c_0(2C_s)^5\left(\frac{4kT}{q}\right)^4(\beta\tan\beta)^5\frac{d}{d\beta}(\beta\tan\beta)\right]d\beta$$

$$= \mu\frac{4C_s W}{L}\left(\frac{2kT}{q}\right)^2 x$$

$$x\left[\beta\tan\beta - \frac{\beta^2}{2} + a_0 C_s \beta^2\tan^2\beta + 3b_0(2C_s)^3\left(\frac{2kT}{q}\right)^2\beta^4\tan^4\beta + \frac{5}{6}c_0(2C_s)^5\left(\frac{4kT}{q}\right)^4\beta^6\tan^6\beta\right]_{\beta_d}^{\beta_s} \quad (7)$$

As expected, (7) reduces to the current in a DG-MOSFET [9] after the identification $a_0=1/C_{ox}$, $b_0=c_0=0$. To compute the drain current we define the following two functions representing the RHS of (5) and (7):

$$f(\beta) = \ln(\beta) - \ln(\cos\beta) + a_0(2C_s)\beta\tan\beta + b_0(2C_s)^3\left(\frac{4kT}{q}\right)^2\beta^3\tan^3\beta + c_0(2C_s)^5\left(\frac{4kT}{q}\right)^4\beta^5\tan^5\beta \quad (8)$$

$$g(\beta) = \beta\tan\beta - \frac{\beta^2}{2} + a_0C_s\beta^2\tan^2\beta + 3b_0(2C_s)^3\left(\frac{2kT}{q}\right)^2\beta^4\tan^4\beta + \frac{5}{6}c_0(2C_s)^5\left(\frac{4kT}{q}\right)^4\beta^6\tan^6\beta \quad (9)$$

The range of $\beta$ is $0<\beta<\pi/2$. For given $V_g$ and $V_{ds}$, $\beta_s$ and $\beta_d$ are calculated from the conditions $f(\beta_s)=(q/2kT)(V_g-V_0)$ and $f(\beta_d)=(q/2kT)(V_g-V_0-V_{ds})$, where

$$V_0 = \Delta\varphi + \frac{2kT}{q}\ln\left(\frac{2}{t_s}\sqrt{\frac{2\varepsilon_s kT}{q^2 n_i}}\right) \quad (10)$$

From (9), the drain current $I_{ds} \propto g(\beta_s) - g(\beta_d)$ can be easily computed. Finding $\beta_s$ and $\beta_d$ can be geometrically interpreted as the intersection of $f(\beta)$ with the load line $f(\beta_s)$ and $f(\beta_d)$, respectively. Fig. 2 illustrates this situation using SBT as the ferroelectric material, characterized by the Landau parameters: $a=-1.3\cdot10^8$, $b=1.3\cdot10^{10}$, $c=0$ (SI units) [5]. Depending on the specific device geometry and ferroelectric material, $f(\beta)$ could be monotonous over the whole range of $\beta$, yielding a single-valued solution for both $\beta_s$ and $\beta_d$, or, on the contrary, could exhibit a non-monotonous behavior, yielding a multi-valued solution for $\beta_s$ or both $\beta_s$ and $\beta_d$. The non-monotonous case is representative of hysteretic behavior. Combination of device geometries and ferroelectric materials conducting to hysteresis should be avoided for conventional CMOS-like operation. Next, we examine the step-up conversion capability. Fig. 3 shows a plot of $\varphi_s$ versus $V_g$ of the MFS structure for several $t_f$ values. As a general rule, high-$t_f$ values give rise to hysteretic behavior. Reducing $t_f$, hysteresis disappears and gain (G) >> 1 can be reached, where G is defined as $dQ/dV_g$ (see Fig. 3: inset). For low $t_f$ values, the step-up conversion capability disappears. As a consequence, an important part of the device design is to properly tune $t_f$. Remarkably, *the signature of operation in the negative*

*capacitance region is a single-valued and peaked $C_g$-$V_g$ characteristic* that could be easily determined using an impedance analyzer (Fig. 4). A sharp peak is indicative of a huge step-up conversion factor. For comparison, the $C_g$-$V_g$ curve of the equivalent MIS structure is also shown. The step-up conversion capability can be used as a mechanism to obtain steeper subthreshold transistors. Fig. 5 shows a plot of the transfer characteristic at room temperature for a negative capacitance FET with $t_s$=5 nm, assuming a SBT ferroelectric oxide. The work-function difference has been tailored to $\Delta\phi$=-0.14 eV to provide an OFF-state current $I_{OFF} \sim 3\cdot10^{-2}$ mA/μm, representative of a low stand-by power transistor for the present 2010 node [2]. The ferroelectric oxide thickness that maximizes $G$ has been found to be $t_f$=20 nm. Note that $G$>1 in the MFS structure automatically translates into $S$<60 mV/decade in the transistor transfer characteristic. This behavior is only observed in a limited range of $V_g$. For comparison purposes, the transfer characteristic of an equivalent positive capacitance FET with $C_{ox}$=1/$|a_0|$ is plotted. Remarkably, the step-up conversion property can be used in CMOS logic to reduce the $V_{dd}$ bias voltage. For instance, if the ON-state current is fixed to $I_{ON} \sim 600$ mA/μm, representative of the 2010 node, a reduction of $V_{dd} \sim 150$ mV can be achieved. This property can also be used to increase the $I_{ON}/I_{OFF}$ figure-of-merit for a prescribed $V_{dd}$. In the finite range where step-up conversion occurs, $I_{ON}$ for a negative capacitance FET is superior to the equivalent positive capacitance FET, as shown in Fig. 6, where the output characteristics are compared.

In practice, MFS structures are very difficult to process, and an insulating buffer layer is needed between the ferroelectric and the semiconductor to properly match their lattice parameters, avoid interdifussion problems between both materials and rule out chemical reactions that degrade the properties of the ferroelectric oxide, the underlying semiconductor, or both, leading to electrically active defects at the semiconductor/interface [10]. The resulting MFIS structure can also be the building block of a negative capacitance FET, but an effective $C_f$ value must be considered to take into account the additional capacitance introduced by the buffer layer in series with the ferroelectric oxide. Even, in the case that a MFS gate stack could be properly engineered, the presence of a non-switchable insulating dead layer in the ferroelectric [11] would convert the MFS gate stack into a MFIS gate stack in practice. The mentioned effect can be accounted for simply by using an effective $a_0$, $a_0^{eff} = a_0 + (1/C_b)$, where $C_b$ represents the buffer layer capacitance including the eventual presence of the dead layer The remaining parameters $b_0$, $c_0$ of the ferroelectric oxide keep their values unchanged.

In conclusion, a physics-based analytical surface potential and drain current model for long-channel negative capacitance FETs is derived from the phenomenological Landau-Ginzburg-Devonshire ferroelectric theory, the Poisson's equation, and the current continuity equation. A typical perovskite SBT ferroelectric has been considered to quantitatively illustrate the model outcomes, but much more can be achieved in terms of gain and corresponding $V_g$ range using properly engineered ferroelectric materials [10,12]. The presented model has been most discussed for operation in the negative capacitance region of the ferroelectric looking for low-power switching applications, but the model formulation is more general and additionally describes the hysteretic behavior and is therefore useful as a surface potential and drain current model for non-volatile memories based on ferroelectric field-effect transistors (Fe-FET).

## ACKNOWLEDGMENTS

This work was supported by the Ministerio de Ciencia e Innovación under projects Explora TEC2008-01883-E/TEC and EUROSOI+. DJ acknowledges stimulating discussions with F. Campabadal, J. M. Rafí, F. Sánchez, H. Taniguchi, S-M. Yoon, I. Ishiwara, A. Cano, A. P. Levanyuk, and M. Kindelan.

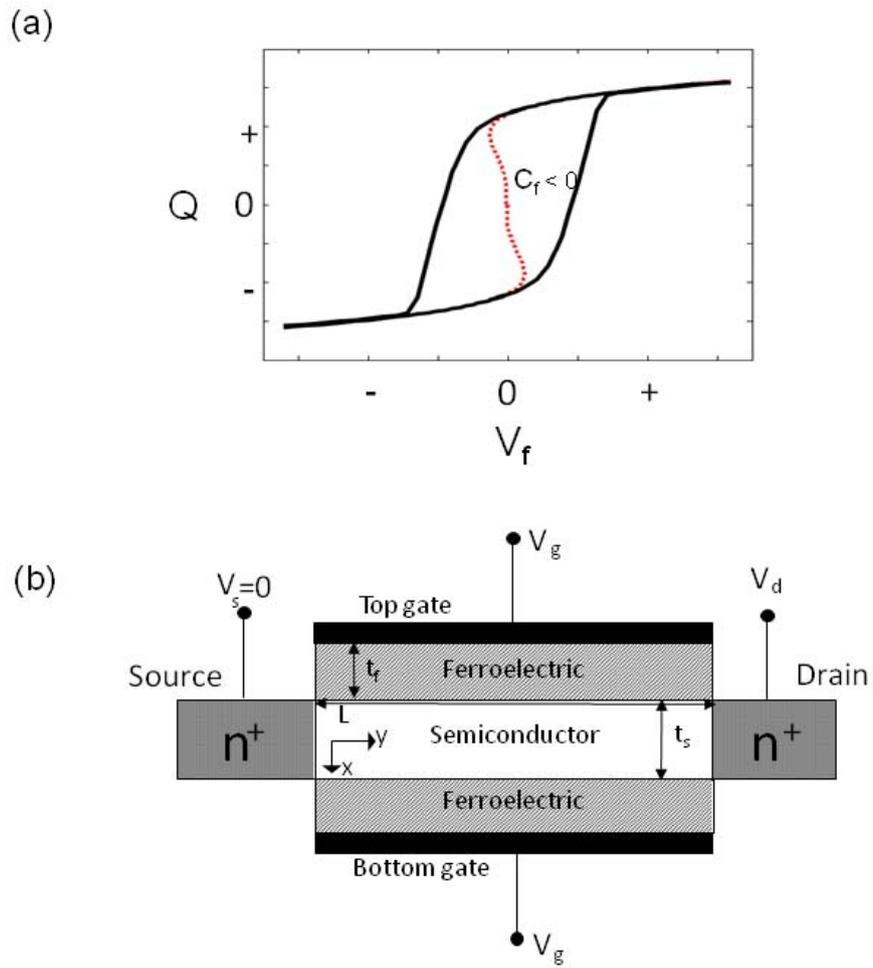

**Figure 1**. (a) Charge versus applied voltage in a ferroelectric oxide film. The solid line represents the typical hysteretic behaviour observed in experiments. The dotted line represents the negative capacitance region. (b) Cross-sectional view of a double-gate negative capacitance FET.

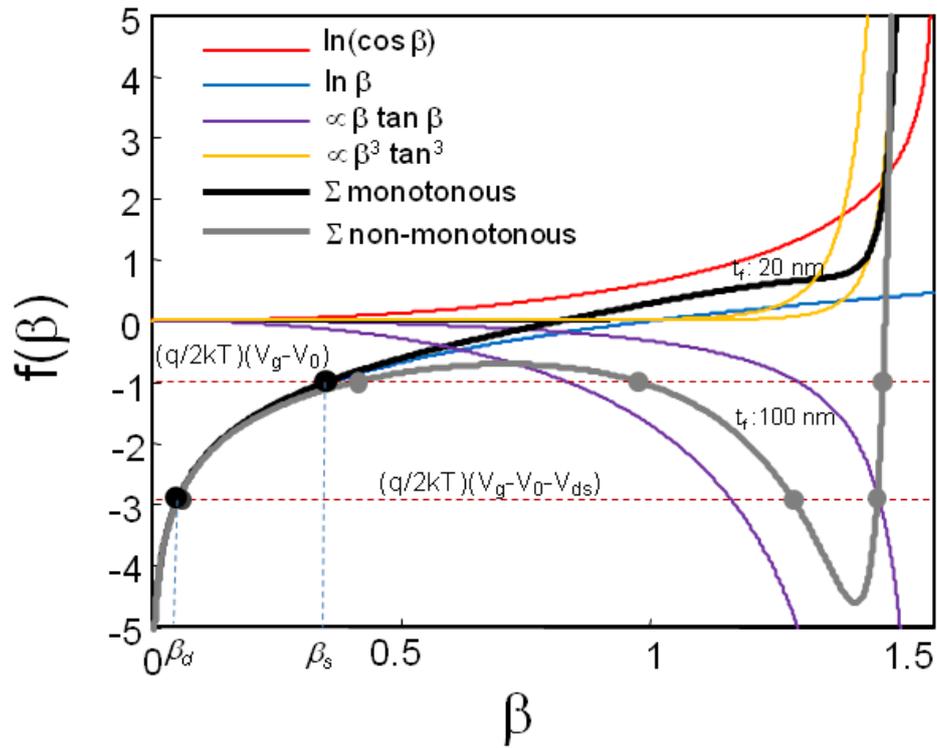

**Figure 2**. Determination of $\beta_s$, $\beta_s$ can be geometrically interpreted as the intersection of $f(\beta)$ with the load line $(q/2kT)(V_g-V_0)$ and $(q/2kT)(V_g-V_0-V_{ds})$ respectively. The curve $f(\beta)$ has been decomposed in the terms shown in (8)

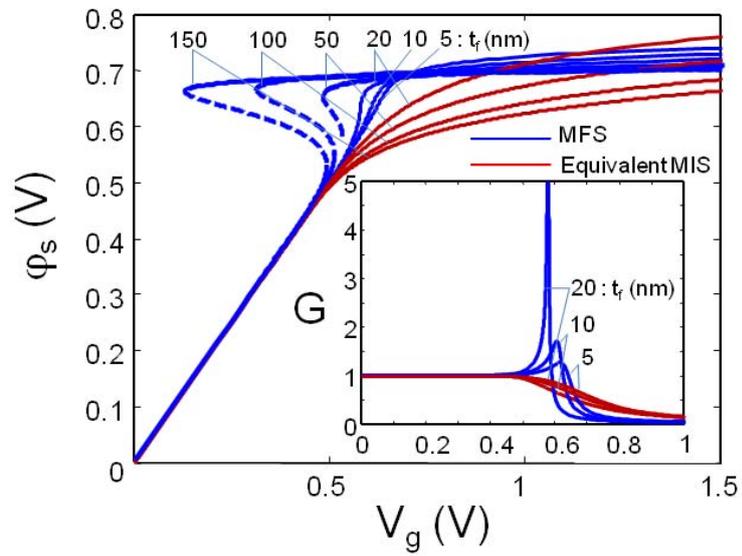

**Figure 3**. Step-up conversion capability of the MFS structure as a function of the ferroelectric film thickness (blue line). Hysteretic behaviour arises for $t_f$ values greater than 20 nm. The dashed blue line indicates the unstable region. For comparison, the surface potential of the equivalent MIS structure is shown in red line. The inset shows the expected gain versus gate voltage.

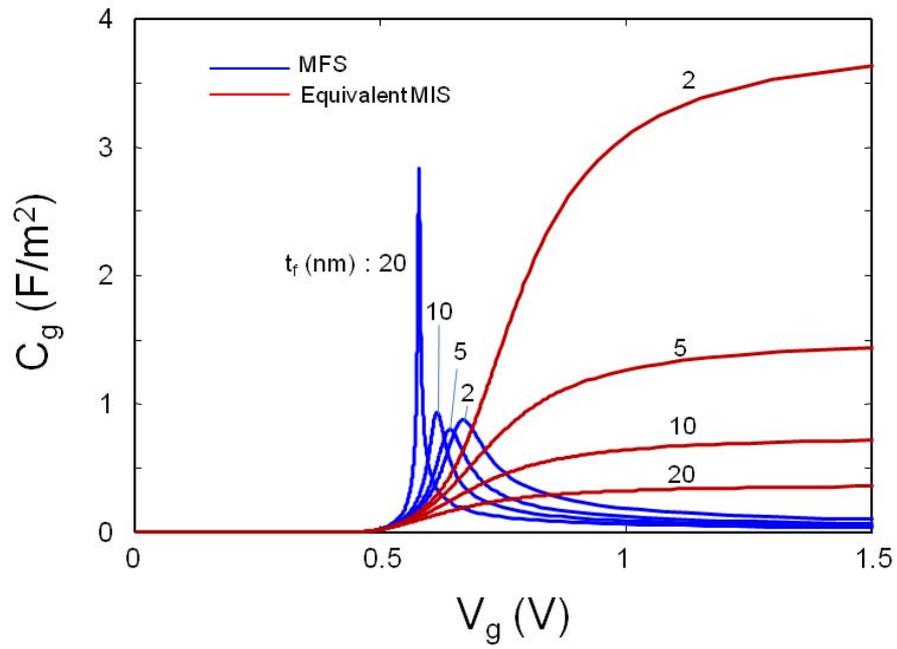

**Figure 4**. The signature of operation in the negative capacitance region is a single-valued and peaked $C_g$-$V_g$ characteristic (blue line). For comparison, the gate capacitance of the equivalent MIS structure is shown in red line.

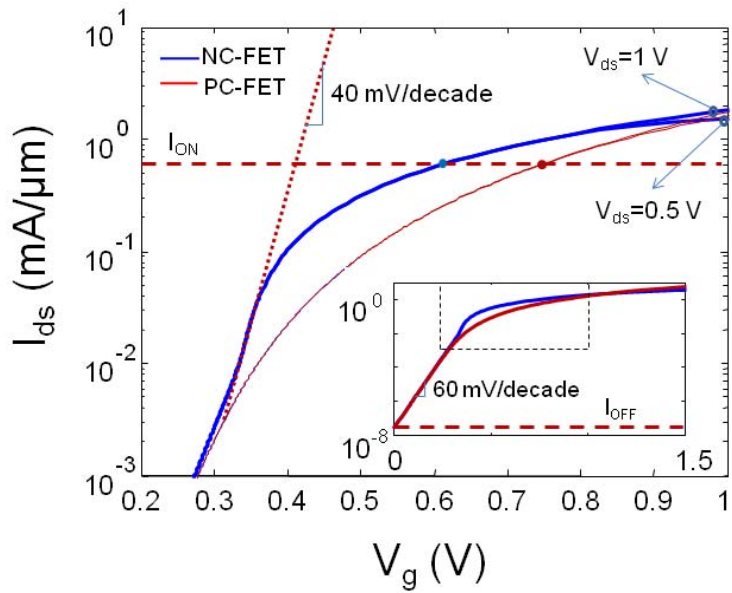

**Figure 5**. Comparison between transfer characteristics of a negative capacitance (NC) FET (blue line) and the equivalent positive capacitance (PC) FET (red line) for a fixed OFF-state current. Room temperature is considered. The transfer characteristic of the negative capacitance FET shows a finite range of $V_g$ where $S<60$ mV/decade is reached. This property can be used to reduce the bias voltage about 150 mV for a fixed ON-state current of 600 mA/μm. Inset: expanded view of the transfer characteristics.

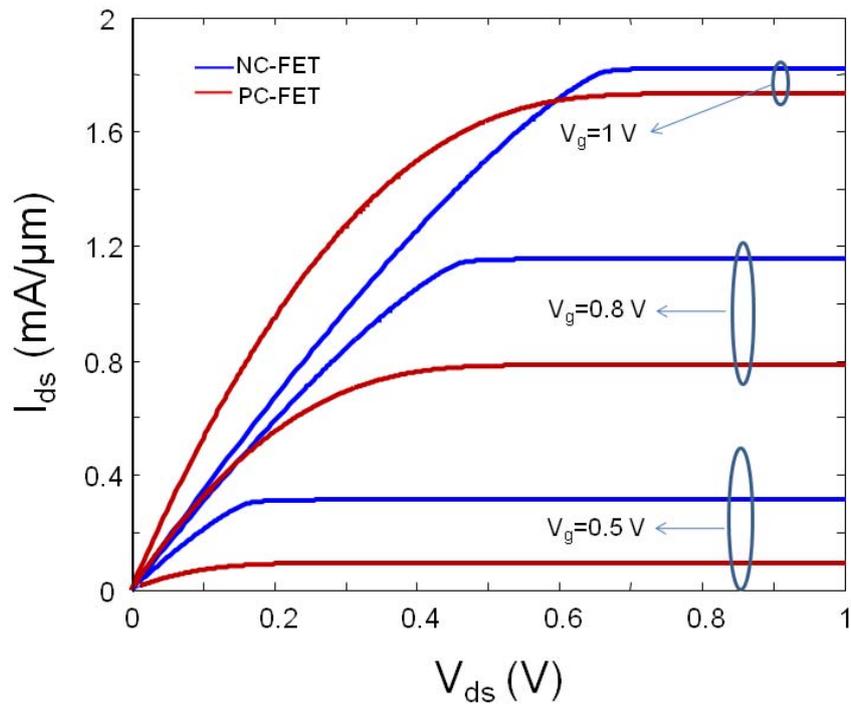

**Figure 6**. Comparison between the output characteristics of a negative capacitance (NC) FET and the equivalent positive capacitance (PC) FET at different gate voltages. Room temperature is considered. In the analyzed range of gate voltages, relevant for low-power switching applications, the negative capacitance FET exhibits a superior ON-state current.